# A Complete Approach to Time Varying Linear Systems

Douglas R. Frey, *Fellow, IEEE*

**Abstract:** This paper presents a unifying theory of Linear second order systems that allows time-varying and time invariant systems to be treated in the same way for the first time. In the process, a transformation is given that diagonalizes an arbitrary time varying state matrix in a spectrum invariant way. A canonical form for the fundamental matrix is given that depends on dynamic eigenvalues and related eigenvectors dependent upon the Riccati Characteristic Equation for the system, which intuitively generalizes the standard characteristic equation for time invariant systems. The technique is shown by examples to give a unified approach to the solutions of time invariant, time-varying, and periodic systems.

Index terms–Active filters, Control systems, Dynamical systems, Eigenvalues and eigenfunctions, Engineering education, Modulation, State-space methods, Time-varying systems[1].

## I. Introduction

The study of linear time varying systems has received much attention in the literature, both in regard to research and practice. This is primarily because many systems of interest fall outside the scope of well understood linear time invariant system analysis. For example, weakly nonlinear systems such as oscillators and communications electronics require careful approximation to adequately characterize their behavior [18-20]. The work in [7,8] has shown how a time varying approach may improve the design and analysis of modulators, complex filters, and super-regenerative receivers. Many areas of design, from audio [22] and medical electronics [23] to time varying filter banks [21], require attention to time varying system behavior. Clearly, broadly used adaptive filters and many controls are time varying systems whose analysis and design would benefit from a complete time varying systems theory. Unfortunately, a well articulated theory of linear time varying systems is not available, despite a considerable wealth of literature [9,11-13]. Nevertheless, various studies, including periodic systems [13], transform analysis [14,24,25] and signal analysis [26,27], all exploit ideas related to time varying system analysis. This paper attempts to bridge the gap between the well understood theory of linear time invariant systems and the poorly understood behavior of linear time varying systems by introducing a unifying analysis for both for the first time. While not covered here, its extension to discrete time should be straightforward. It is expected that this work will impact not only the work of scientists and engineers, but the teaching of general linear systems as well.

Linear systems are generally divided into 2 classes– namely, linear time-invariant, or LTI, and linear time-varying, or LTV systems. All LTI systems have known explicit solutions. Some LTV systems also admit to explicit solutions; however, the approach to solving these systems is typically quite different than that for LTI systems. Transforms are not reasonably available for LTV systems, for example. Nevertheless, there has been a continuing desire over the years to analyze LTV systems using techniques applicable to LTI systems. The work in [9-11] are examples of some of the serious attempts. This paper presents a comprehensive approach to giving LTI and LTV systems a common analysis technique. It is granted that classical techniques are still extremely valuable, but they do not extend naturally to LTV systems.

In what follows, we focus on general second-order LTV systems, taking LTI systems as a subset. This allows the paradigm to be offered in a relatively simple, yet conceptually complete, way. We conclude with a discussion of the extension to higher order systems. The analysis provided is applicable to all second order systems, and may be easily simplified in dealing with LTI systems. It is useful to begin by highlighting the issues that motivate this work. In LTI systems one focuses on steady state and transient behavior, which inevitably leads to the ideas of fundamental, or core, modes, characterized by poles, or eigenvalues in the state space context, and stability. The duality between time domain and frequency domain analysis in LTI systems is elegant and extremely useful. Unfortunately, it precipitates an underlying thought process that is not conducive for the study of LTV systems, which may only be studied productively in the time domain. It is also unfortunate that the linear algebra applied to LTI state space systems does not generally extend in an intuitive way to LTV systems. For instance, the eigenvalues of a constant state matrix are naturally related to the fundamental modes of the associated LTI system through integration and exponentiation. Allowing the same steps in computing eigenvalues for constant matrices to be used for time varying matrices does not result in dynamic eigenvalues pertinent to the associated LTV system fundamental modes. This is because one may apply any linear time invariant transformation to a constant matrix with the assurance that the eigenvalues of the matrix are invariant, such as the transformation that diagonalizes it. This is not the case for time varying matrices, since diagonalization cannot be

The author is with the ECE Dept., Lehigh University, Bethlehem, PA 18015
(e-mail: drf3@lehigh.edu)

achieved via a time invariant transformation. It is intriguing to note that LTV systems possess a state transition matrix that is used in the same convolution integral as LTI systems to produce the zero state response. This further suggests that a straightforward generalization of LTI analysis may be applied to LTV systems. Despite good attempts [e.g.3,5], no researcher has effectively shown how to do this. The work here demonstrates that an intuitive and productive generalization does exist.

This paper is organized in the following way. The Introduction continues with a summary of an approach to solving for fundamental modes in second order LTV systems in a context appropriate to the ensuing discussion. The key to finding the fundamental modes, inherently related to dynamic eigenvalues, is to solve a Riccati differential equation related to a given LTV system. We call this Riccati equation the Riccati Characteristic Equation, or RCE. While this idea has been proposed elsewhere [2-4], no one has shown the connection of the RCE to solutions and the fundamental matrix in the compelling and unifying way here. For example, as proven in the companion paper [1], the structure of solutions to the RCE is exactly the same as those of the classic characteristic equation. Section II continues by introducing a new approach to transforming state equations that for the first time preserves the spectrum of the state matrix in the process of finding its diagonal form. This process is shown to appropriately identify the fundamental matrix as the product of a dynamic eigenvector matrix and an exponential matrix determined by the dynamic eigenvalues of the system. The simply specified form of the fundamental matrix, coupled with its connection to the Riccati Characteristic Equation (RCE), allows a unique straightforward way to anticipate the Floquet theorem for LTV second order systems. Section III uses the section II results along with the RCE to completely specify the solutions to arbitrary LTV systems, including how arbitrary boundary conditions in the time domain translate to a single necessary boundary condition in the Riccati domain, where the RCE is solved. Section IV then offers a number of examples fully articulating the use of the current approach in finding the solution of LTI and LTV systems, including a periodic LTV example. New perspectives on system response are given. The final section ends the paper with a summary of the methodology presented here including insights into the extension of this paradigm to higher order systems.

Let us begin by looking at second order time varying linear systems in the time domain. In practice, the considered systems are often characterized using either scalar second order differential equations or second order dynamical equations. In the scalar differential equation form we have the following, where it has been assumed that any coefficient of the second order derivative term has isolated zeros and may be divided:

$$\ddot{y}(t) + r_1(t)\dot{y}(t) + r_0(t)y(t) = u(t) \tag{1}$$

The input, u(t), here captures a cumulative input comprising a sum of a scaled actual system input and its derivatives. A companion form for this differential equation is given by these state equations, where $y(t) = x_1(t)$.

$$\begin{bmatrix} \dot{x}_1(t) \\ \dot{x}_2(t) \end{bmatrix} = \begin{bmatrix} 0 & 1 \\ -r_0(t) & -r_1(t) \end{bmatrix} \begin{bmatrix} x_1(t) \\ x_2(t) \end{bmatrix} + \begin{bmatrix} 0 \\ 1 \end{bmatrix} u(t) \tag{2}$$

The system described by these equations is a special case of a general second order time varying dynamical equation. Hence, we consider the broad class of time varying second order dynamical equations without loss of generality. Clearly, the crux of the problem in solving such systems is that of solving the state equations. Therefore, we focus on that core problem in addressing LTV analysis. Specifically, we consider general second order linear systems which are described by the state equations shown below.

$$\begin{bmatrix} \dot{x}_1(t) \\ \dot{x}_2(t) \end{bmatrix} = \begin{bmatrix} a_{11}(t) & a_{12}(t) \\ a_{21}(t) & a_{22}(t) \end{bmatrix} \begin{bmatrix} x_1(t) \\ x_2(t) \end{bmatrix} + \begin{bmatrix} u_1(t) \\ u_2(t) \end{bmatrix} \tag{3}$$

It is well known that the key to solving this non-autonomous system lies in solving the autonomous system where the input vector, $\bar{u}(t)$, is set to zero. In light of that, we focus on solutions to the autonomous state equations. To that end, let us suppose that any solution to the system comprises at most a linear combination of core (fundamental) solutions, each of which may be written in terms of a dynamic eigenvalue, for example $\lambda(t)$, as shown here, where y(t) represents the respective core function.

$$y(t) = e^{\int \lambda(t)dt} = e^{\varphi(t)} \; ; \; \varphi(t) = \int \lambda(t)dt \tag{4}$$

This is completely consistent with classical LTI system analysis where $\lambda(t)$ is a constant, simply called an eigenvalue. It is therefore appropriate to look for all dynamic eigenvalues that pertain to a given system.

Suppose that we look at the solution for $x_1(t)$ in the autonomous version of (3). Clearly, the solution of just one of the state variables enables the solution of the other, with the solution of at most a first order linear differential equation. It should be mentioned that in a time varying system the dynamic eigenvalues found by focusing on $x_2(t)$ will appear different than those for $x_1(t)$ since, as will be shown, the core solutions will be those of $x_1(t)$ multiplied by a time function. Consequently, we are declaring that the system core modes are referenced to a particular system variable. To that end, let us suppose that $x_1(t)$ is given by the exponential in (4), where $\varphi(t) = \int a_{12}(t)z(t)dt$. This particular guess at the solution yields a constraint on the unknown function, z(t). Specifically, where all parameters are assumed to be time-varying and $a_{ij}(t)$ is the ij-th element of the state matrix, we have,

$$\dot{x}_1 = za_{12}e^{\varphi} \Rightarrow x_2 = (z-a)e^{\varphi}, a = \frac{a_{11}}{a_{12}}$$
$$\dot{x}_2 = [(\dot{z}-\dot{a}) + za_{12}(z-a)]e^{\varphi} = a_{21}x_1 + a_{22}x_2$$
$$\dot{z} = -a_{12}z^2 + (a_{22} + aa_{12})z + \dot{a} - a_{22}a + a_{21}$$
$$\dot{z} = -a_{12}(z-\eta)^2 + a_{12}\eta^2 + \dot{a} - a_{22}a + a_{21} \quad (5)$$
$$\dot{z} - \dot{\eta} = -a_{12}(z-\eta)^2 + \dot{\alpha} + a_{12}\alpha^2 + a_{21}$$
$$\alpha = \frac{a_{11} - a_{22}}{2a_{12}}; \eta = \frac{a_{11} + a_{22}}{2a_{12}}; \nu = z - \eta$$
$$\therefore \dot{\nu} = -a_{12}\nu^2 + \omega_{02}; \omega_{02} = \dot{\alpha} + a_{12}\alpha^2 + a_{21}$$

Observe that z(t) is a shifted version of ν(t), where ν(t) is a solution to a compact Riccati differential equation. From here on the reduced Riccati differential equation for ν(t) in (5) shall be called the Riccati characteristic equation, or RCE, for the system. It is well understood that the Riccati equation has more than one solution [e.g.14]. Consequently, the multiple solutions of the respective Riccati equations for z(t) and ν(t) are identical except for an offset–namely, η(t). The Riccati characteristic equation is the characteristic equation for all linear second order systems, since its solutions, with an offset, yield the dynamic eigenvalues via multiplication by $a_{12}(t)$. To see this, consider the LTI case where the coefficients of this equation are constant. Recognizing that α is a constant, there is a pair of constant solutions for both ν(t) and z(t). Specifically, we may write the solutions for ν and $\lambda = a_{12} z$.

$$\nu = \pm\sqrt{\frac{\omega_{02}}{a_{12}}} = \sqrt{\frac{(a_{11}-a_{22})^2}{4a_{12}^2} + \frac{a_{21}}{a_{12}}} \quad (6)$$
$$\lambda = b \pm \sqrt{b^2 - (a_{11}a_{22} - a_{12}a_{21})}; b = \frac{a_{11}+a_{22}}{2}$$

The final result for λ is in fact that found by solving the classical quadratic characteristic equation for the system to obtain eigenvalues of the matrix A (a constant matrix in this case).

An important aspect of the solution for $a_{12}z$ in this classic LTI case is that it provides a pair of eigenvalues which may both be real or a pair of complex conjugates. When these eigenvalues are integrated and exponentiated, we obtain a pair of core solutions to the homogeneous linear system. All linear combinations of these core solutions qualify as a composite solution to that system. In the case of complex conjugate eigenvalues, the core solutions are themselves complex conjugates which must be paired appropriately to yield real solutions to the system. It is noteworthy that the Riccati equation above always possesses purely real solutions, even in the LTI case, which will in general provide time varying functions for $a_{12}z$ that, when integrated and exponentiated, are the real (not complex) solutions for the linear system. This will be shown later.

Perhaps the most important aspect of the use of the Riccati characteristic equation for all systems and its correspondance to the classic LTI case characteristic equation is that the solutions to the Riccati equation may be examined in pairs as well. As shown in [1] the RCE always has primitive solutions that may be put in the form, $\nu_R(t) \pm \nu_I(t)$, where $\nu_R(t)$ is the real component and the intrinsic component, $\nu_I(t)$, is either real or purely imaginary. This offers a uniformity to the analysis across the full spectrum of LTV systems.

**II. Optimal Transformations for the Solution of Systems**

Given a system described by the state equations of (3), there is a general approach to the transformation of these systems that yields dynamic eigenvalues, eigenvectors, and fundamental matrices. This approach maintains the core solutions in the process, for the first time showing how to diagonalize the state matrix in a fundamentally meaningful way. We begin the discussion by considering the effect of state space transformations on the state matrix and the associated solutions of the system of state equations.

Given a system described in (3), we may transform the state space with a nonsingular matrix, M(t), yielding a new set of state equations with a transformed state matrix. This is not a new idea [4,5,9], but it is explored differently here. A well chosen matrix transformation will yield new state equations that may admit to an analytic solution, or a desired form for the state matrix. This process is shown below.

$$\dot{\overline{x}}(t) = A(t)\overline{x}(t) + \overline{u}(t); let \ \overline{w}(t) = M(t)\overline{x}(t)$$
$$\dot{\overline{w}}(t) = [M(t)A(t) + \dot{M}(t)]\overline{x}(t) + M(t)\overline{u}(t)$$
$$= A_M(t)\overline{w}(t) + M(t)\overline{u}(t) \quad (7)$$
$$A_M(t) = M(t)A(t)M^{-1}(t) + \dot{M}(t)M^{-1}(t)$$

The new state matrix, $A_M(t)$, will in general have different dynamic eigenvalues than the original state matrix, A(t). This translates to the fact that the core solutions of the new state equations will in general differ from those of the original system. This may or may not be a good thing. One very good case occurs when the new state matrix is transformed to a matrix of zeros. In this case, the transformed system is trivial to solve. Then, using an inverse transformation, one may easily find the solutions to the original system, as shown next. Setting $A_M(t) = 0$, we have,

$$\int_0^t \dot{\overline{w}}(\tau)d\tau = \overline{w}(t) - \overline{w}(0) = \int_0^t M(\tau)\overline{u}(\tau)d\tau$$

$$\overline{w}(t) = M(0)\overline{x}(0) + \int_0^t M(\tau)\overline{u}(\tau)d\tau \tag{8}$$

$$\overline{x}(t) = \phi(t)\overline{x}(0) + \phi(t)\int_0^t \phi^{-1}(t)\overline{u}(\tau)d\tau$$

In (8) we have defined $\phi(t) = M^{-1}(t)M(0)$. Observe that knowledge of $\phi(t)$ solves the system and is, in fact, the classical state transition matrix.

M(t) is clearly at the heart of the problem in solving the system, since it can transform the state matrix to a form which will allow a solution to the original system. However, in general M(t) is not a matrix that is easily found. Furthermore, M(t) is not obviously found by approximation. This leads to a related approach which has engineering significance. This approach requires multiple simple transformations, but offers much more insight. To begin, suppose that the following transformation–that is, $M_0(t)$– is used as a first step. Then the new state matrix, $A_0(t)$, is found as shown, where all scalar quantities are time varying.

$$M_0(t) = \begin{bmatrix} 1 & 0 \\ \alpha & 1 \end{bmatrix}; M_0^{-1}(t) = \begin{bmatrix} 1 & 0 \\ -\alpha & 1 \end{bmatrix}$$

$$A_0(t) = M_0(t)A(t)M_0^{-1}(t) + \dot{M}_0(t)M_0^{-1}(t)$$

$$= \begin{bmatrix} \sigma_1 & \omega_{01} \\ \omega_{02} & \sigma_2 \end{bmatrix}; \omega_{01} = a_{12} \tag{9}$$

$$\sigma_1 = a_{11} - \alpha a_{12}; \sigma_2 = a_{22} + \alpha a_{12}$$

$$\omega_{02} = a_{21} + \alpha(a_{11} - a_{22}) - \alpha^2 a_{12} + \dot{\alpha}$$

Observe that if $\alpha(t)$ is chosen to be the solution to the Riccati equation, where $\omega_{02}(t) = 0$, then the state matrix becomes triangular, enabling an explicit solution for the system [4]. On the other hand, the state matrix may be put into the form of (9) with equal diagonal elements–that is, $\sigma_1(t) = \sigma_2(t) = \sigma_0(t)$. The equality of the diagonal elements requires the constraint on $\alpha(t)$, shown below, yielding the following results, where $\omega_{02}(t)$ is not likely to equal $\omega_{01}(t)$ or zero.

$$\sigma_0(t) = \frac{a_{11}(t) + a_{22}(t)}{2}; \alpha(t) = \frac{a_{11}(t) - a_{22}(t)}{2a_{12}(t)} \tag{10}$$

$$\omega_{02}(t) = a_{21}(t) + a_{12}(t)\alpha^2(t) + \dot{\alpha}(t)$$

For reasons related to engineering applications, a state matrix having this form of $A_0(t)$ will be said to be in the *modulation form*. An important benefit of this transformation to the modulation form is that the core solutions of the transformed homogeneous system are the same as the original. This is proven next, where $\sigma_1(t) = \sigma_2(t) = \sigma_0(t) = (a_{11}(t) + a_{22}(t))/2 =$ $\omega_{01}(t) \eta(t)$, and $\omega_{01}(t) = a_{12}(t)$.

$$\dot{x}_1 = \sigma_0 e^\varphi + \omega_{01}x_2 \Rightarrow x_2 = (z - \eta)e^\varphi, \eta = \frac{\sigma_0}{\omega_{01}}$$

$$\dot{x}_2 = [(\dot{z} - \dot{\eta}) + z\omega_{01}(z - \eta)]e^\varphi = \omega_{02}x_1 + \sigma_0 x_2 \tag{11}$$

$$\dot{z} = -\omega_{01}(z - \eta)^2 + \omega_{01}\eta^2 + \dot{\eta} - \omega_{01}\eta^2 + \omega_{02}$$

$$\dot{v} = \dot{z} - \dot{\eta} = -\omega_{01}v^2 + \omega_{02}$$

Given $\omega_{01}(t) = a_{12}(t)$, and comparing (5) to (11), where the definitions for $\omega_{02}(t)$ and $\eta(t)$ are the same as in (5),(10), and(11), clearly both systems have the same dynamic eigenvalues. Specifically, the RCE for v(t) and its relation to z(t) are identical in both cases.

Now we propose a transformation matrix, P(t), that transforms the state matrix, $A_0(t)$, to a final modulation form, $A_f(t)$, where the off-diagonal elements are equal in magnitude. We refer to this form for $A_f(t)$ as the *symmetric modulation form*. Specifically,

$$A_f(t) = P(t)A_0(t)P^{-1}(t) + \dot{P}(t)P^{-1}(t)$$

$$= \begin{bmatrix} \sigma_f(t) & \omega_f(t) \\ \omega_f(t) & \sigma_f(t) \end{bmatrix} \tag{12}$$

$$P(t) = \begin{bmatrix} 1 & 0 \\ -q(t) & p(t) \end{bmatrix}; q(t) = \frac{\dot{p}(t)}{2\omega_{01}(t)}$$

The function, p(t), is the solution of a linear first order differential equation that depends on $\omega_{01}(t)$ and v(t) = z(t) - $\eta$(t), where v(t) is the solution to the simplified RCE specified in (5) and (11). The equation for p(t) is given below.

$$\dot{p}(t) = 2\omega_{01}(t)v(t)p(t) - 2\omega_{01}(t) \tag{13}$$

It is shown in Appendix 1 that the matrix, P(t), transforms the state matrix, $A_0(t)$, to the symmetric modulation form given by $A_f(t)$, shown in (12), where,

$$\omega_f(t) = \frac{\omega_{01}(t)}{p(t)}; \sigma_f(t) = \frac{\dot{p}(t)}{2p(t)} + \sigma_0(t) \tag{14}$$

Observe that the symmetric modulation form of $A_f(t)$ makes its dynamic eigenvalues, $\lambda_{1,2}(t)$, equal to $\sigma_f(t) \pm \omega_f(t)$. It is now proven that these yield the same core solutions as the original system state matrix, A(t), where (13) is exploited.

$$\sigma_0(t) + \frac{\dot{p}(t)}{2p(t)} = \sigma_0(t) + \omega_{01}(t)v(t) - \frac{\omega_{01}(t)}{p(t)}$$

$$\lambda(t) = \omega_{01}(t)v(t) + \sigma_0(t) - \frac{\omega_{01}(t)}{p(t)} \pm \frac{\omega_{01}(t)}{p(t)} \tag{15}$$

$$z(t) = \frac{\lambda(t)}{\omega_{01}(t)} = v(t) + \eta(t) - \frac{1}{p(t)} \pm \frac{1}{p(t)}$$

Clearly, the first (the plus option for ±) solution for z(t) matches the results in (5) and (11), where v(t) solves the RCE. It can be shown [1] that the equation for p(t) guarantees that v(t) - 2/p(t) also solves the RCE with v(t) as a solution; hence, the second solution will also lead to a core solution for the original system. Therefore, the eigenvalues of the state matrix are invariant to the transformation, P(t)M₀(t).

The symmetric modulation form of the state matrix is valuable since the corresponding autonomous system may be readily solved for the state transition matrix, since $A_f(t)$ commutes with its integral. However, given the symmetric modulation form of $A_f(t)$, it is a trivial matter to transform this to its diagonal form, $A_{fd}(t)$, with a matrix, D, as shown next. Since D is a constant matrix, it is clear that the eigenvalues of the state matrix are invariant to the transformation, $M_{fd}(t) = DP(t)M_0(t)$.

$$D = \begin{bmatrix} 1 & 1 \\ 1 & -1 \end{bmatrix} ; \; A_{fd}(t) = DA_f(t)D^{-1}$$

$$A_{fd}(t) = \begin{bmatrix} \sigma_f(t) + \omega_f(t) & 0 \\ 0 & \sigma_f(t) - \omega_f(t) \end{bmatrix} = \begin{bmatrix} \lambda_1(t) & 0 \\ 0 & \lambda_2(t) \end{bmatrix} \quad (16)$$

$$M_{fd}(t) = DP(t)M_0(t) = \begin{bmatrix} 1 - q(t) + \alpha(t)p(t) & p(t) \\ 1 + q(t) - \alpha(t)p(t) & -p(t) \end{bmatrix}$$

This now shows for the first time how to diagonalize a time varying state matrix, preserving its dynamic spectrum. It should be noted that for LTI systems, the matrix that diagonalizes the state matrix, preserving its spectrum, is a constant matrix in general that is the inverse of the matrix of eigenvectors of the state matrix. The eigenvector matrix is not unique since the eigenvectors may be arbitrarily scaled. In addition, as shown by looking at pairs of solutions for the RCE, there is a continuum of pairs of dynamic eigenvalues for a linear system. This is true for LTI systems as well. This aspect will be explored later.

Returning to the set of transformations that diagonalize the state matrix, it is simple to transform this diagonal state matrix to the all zero matrix, using $M_D(t)$, which completes the determination of M(t), and $M^{-1}(t)$, which determines the fundamental matrix for the system.

$$M_D A_{fd}(t) M_D^{-1}(t) + \dot{M}(t) M_D^{-1}(t) = 0$$

$$M_D(t) = \begin{bmatrix} e^{-\int \lambda_1(t)dt} & 0 \\ 0 & e^{-\int \lambda_2(t)dt} \end{bmatrix} \quad (17)$$

Continuing, we may write the final form for M(t), and importantly, $M^{-1}(t)$.

$$M(t) = M_D(t)M_{fd}(t) ; \; M^{-1}(t) = M_{fd}^{-1}(t)M_D^{-1}(t)$$

$$M^{-1}(t) = \begin{bmatrix} 1 - q(t) + \alpha(t)p(t) & p(t) \\ 1 + q(t) - \alpha(t)p(t) & -p(t) \end{bmatrix}^{-1} M_D^{-1}(t)$$

$$= V(t) \begin{bmatrix} e^{\int \lambda_1(t)dt} & 0 \\ 0 & e^{\int \lambda_2(t)dt} \end{bmatrix} \quad (18)$$

$$V(t) = \begin{bmatrix} 1 & 1 \\ \dfrac{q(t)+1}{p(t)} - \alpha(t) & \dfrac{q(t)-1}{p(t)} - \alpha(t) \end{bmatrix}$$

A factor of ½ was left out of the inverse of $M_{fd}(t)$ in the final result since a simple scaling is irrelevant in specifying a fundamental matrix. This result is interesting in that it depends only on the parameter, p(t), and elements of the original state matrix. The function, p(t), of course, depends on the solution to the Ricatti characteristic equation. In fact, the eigenvector matrix, V(t), may be written in terms of the complementary pair of solutions, $v_1(t)$ and $v_2(t)$, to the RCE. Specifically,

$$q(t) = v_1(t)p(t) - 1 \Rightarrow v_1(t) = \frac{q(t)+1}{p(t)}$$

$$v_2(t) = v_1(t) - \frac{2}{p(t)} = \frac{q(t)-1}{p(t)} \quad (19)$$

$$\therefore V(t) = \begin{bmatrix} 1 & 1 \\ v_1(t) - \alpha(t) & v_2(t) - \alpha(t) \end{bmatrix}$$

The final result in (18) shows that $M^{-1}(t)$, the fundamental matrix, is the product of an eigenvector matrix, V(t), and the diagonal core solution matrix. For the LTI case in classical analysis, V(t) is a constant matrix that is found directly using knowledge of the eigenvalues. Using the analysis here for the LTI case, q(t) = 0 and the eigenvector matrix is given as,

$$V = \begin{bmatrix} 1 & 1 \\ \dfrac{1}{p} - \alpha & -\dfrac{1}{p} - \alpha \end{bmatrix} ; \; \alpha = \frac{a_{11} - a_{22}}{2a_{12}} , \; p = \sqrt{\frac{\omega_{01}}{\omega_{02}}} \quad (20)$$

This simple general form for the eigenvector matrix is in itself useful for solving arbitrary LTI systems. The state transition matrix, φ(t), which is the normalized fundamental matrix such that φ(0) is the identity matrix, is shown below. Note that in the LTI case, M(0) = V⁻¹(0).

$$\phi(t) = M^{-1}(t)M(0) = V(t) \begin{bmatrix} e^{\int \lambda_1(t)dt} & 0 \\ 0 & e^{\int \lambda_2(t)dt} \end{bmatrix} M(0) \quad (21)$$

It will be shown below that it is not true that V(t) is a constant matrix for the LTI case in general. This is because, when the state matrix is constant, the associated Riccati characteristic equation still admits a continuum of time varying solutions.

Having stated these results, it is interesting to consider the special case where the state matrix is periodic. Specifically, $A(t + T) = A(t)$, for all t and a given T. In this case the functions, $\omega_{01}(t)$ and $\omega_{02}(t)$ are periodic. Suppose that the solutions of the Riccati characteristic equation are periodic. and therefore the dynamic eigenvalues are periodic. It follows from (10), (19), and [1] that the eigenvector matrix, $V(t)$, as well as the dynamic eigenvalues, $\lambda_1(t)$ and $\lambda_2(t)$, are periodic.

Given periodic dynamic eigenvalues, their respective integrals must be the sum of a zero mean periodic function and a constant times time, where the constant is the respective average value associated with each of the dynamic eigenvalues. Thus we may write the fundamental matrix, as in (18), of the LTV system as follows.

$$\phi(t) = V(t)\begin{bmatrix} e^{p_1(t)} & 0 \\ 0 & e^{p_2(t)} \end{bmatrix}\begin{bmatrix} e^{r_1 t} & 0 \\ 0 & e^{r_2 t} \end{bmatrix}$$

$$\int \lambda_{1,2}(t)dt = p_{1,2}(t) + r_{1,2}t \;;\; r_{1,2} = \frac{1}{T}\int_0^T \lambda_{1,2}(t)dt \quad (22)$$

Now we observe that since the transformation, $M(t)$, yields a zero state matrix, then the transformation, $WM(t)$, where W is a nonsingular constant square matrix also yields a zero state matrix, and therefore $M^{-1}(t)W^{-1}$ must be a fundamental matrix for the system. Hence, we may write,

$$\phi(t) = Q(t)W\begin{bmatrix} e^{r_1 t} & 0 \\ 0 & e^{r_2 t} \end{bmatrix}W^{-1} = Q(t)e^{Rt}$$

$$Q(t) = V(t)\begin{bmatrix} e^{p_1(t)} & 0 \\ 0 & e^{p_2(t)} \end{bmatrix}W^{-1} \;;\; R = W\begin{bmatrix} r_1 & 0 \\ 0 & r_2 \end{bmatrix}W^{-1} \quad (23)$$

Clearly, Q(t) is a periodic matrix, whose explicit form is now given, and R is a constant matrix. This anticipates classical Floquet theory [13], noting that $e^{RT}$ is called the monodromy matrix.

There is, however, no research in the literature that guarantees that the periodicity of $\omega_{01}(t)$ and $\omega_{02}(t)$ ensures a periodic solution to the RCE [e.g.16,17]. Some researchers have shown that periodic solutions will exist given extra conditions on the coefficients [e.g.16,17]. But in light of the work here and the known validity of Floquet theory, we can state for the first time that periodic solutions must exist if $\omega_{01}(t)$ and $\omega_{02}(t)$ are periodic. Suppose that the RCE solutions are aperiodic. Then it follows that $V(t)$ is aperiodic. Further, it follows that the extraction of any constants, $r_i$, from the resulting aperiodic dynamic eigenvalues will result in aperiodic functions, $p_i(t)$, and therefore aperiodic core functions. Given the explicit form of V(t), then it is impossible for Q(t) to be periodic for arbitrary constant choices for W in (23). Hence, the assumption that the RCE has no periodic solutions must be invalid, since Floquet's theorem guarantees that Q(t) must be periodic. Furthermore, given the functional dependance relating the real and intrinsic components, $v_R(t)$ and $v_I(t)$, of the solution to the RCE [1], if one solution to the RCE is periodic, then both complementary solutions of the RCE must be periodic. This is a fact never shown before.

### III. General Solution of LTV Systems

Having set the stage by fully articulating the connection between general LTV systems and their corresponding RCE, we may use the results in [1] to compactly write the solution to general systems. To begin, consider the LTI special case. Considering the primitive solutions to the RCE to be the pair of constants, +a and -a below, the general solution for $v(t)$ [1], and the core solutions, $y(t)$, in the time domain, in the case where $\omega_{01}(t)$ and $\omega_{02}(t)$ are constants is given by,

$$v = 0 \pm a, a = \sqrt{\frac{\omega_{02}}{\omega_{01}}} \Rightarrow \lambda = \sigma_0 + \omega_{01}(0 \pm a)$$

$$y(t) = Ae^{\int \lambda dt} = Ae^{\lambda t} = Ae^{\sigma_0 t \pm \omega_{01} a t}$$

$$v(t) = \begin{cases} a\tanh(\omega_{01} a t - K) \\ a\coth(\omega_{01} a t - K) \end{cases} ; \lambda(t) = \sigma_0 + \omega_{01}v(t) \quad (24)$$

$$\Rightarrow y(t) = \begin{cases} Ae^{\sigma_0 t}\cosh(\omega_{01} a t - K) \\ Ae^{\sigma_0 t}\sinh(\omega_{01} a t - K) \end{cases}$$

The constant, or primitive, solutions apply for a or -a as initial conditions to the RCE. Note that for initial conditions of the RCE in the range (-a, a), the tanh solution will be used, and for the range > a or < -a the coth solution will be used. K is an arbitrary slack variable that allows an arbitrary initial condition for the solution of the RCE. The result is core solutions in the time domain with all possible combinations of the basic core exponentials, and just the familiar single exponentials for the primitive case with K infinite.

These statements apply for the case where a is real. For the imaginary case where $\omega_{01}(t)$ and $\omega_{02}(t)$ are constants of opposite sign, the primitive solutions are $\pm ja$ and we have a pair of complex conjugate exponentials for the primitive case. Allowing arbitrary real initial conditions for the RCE the hyperbolic functions become trigonometric, with the following results.

$$v(t) = \begin{cases} -a\tan(\omega_{01}at - K) \\ a\cot(\omega_{01}at - K) \end{cases} ; \; v_I = ja = \sqrt{\frac{\omega_{02}}{\omega_{01}}}$$

$$y(t) = \begin{cases} Ae^{\sigma_0 t}\cos(\omega_{01}at - K) \\ Ae^{\sigma_0 t}\sin(\omega_{01}at - K) \end{cases} \quad (25)$$

Note that this is precisely the result obtained by replacing a with ja and K with jK in (24) after simplification.

It should be further noted that the core solutions, as in (4), in the case of complex conjugate solutions for the Riccati characteristic equation according to (25) are each basically sinusoidal, which are essentially equivalent allowing an obvious phase shift. This represents the true physical solution to a system with complex conjugate eigenvalues. The solution may be scaled and phase shifted arbitrarily to satisfy initial conditions in the time domain.

This generalization of the perspective in obtaining solutions to LTI systems suggests an intriguing similar perspective in looking at the solution of LTV systems. Recognizing that the given analysis is completely valid for a general LTV system, the fundamental matrix for the original system having A(t) as its state matrix is that shown in (18). In this case we have time varying dynamic eigenvalues and eigenvectors completely specified in section II As shown in [1], dealing with solutions of the RCE, the dynamic eigenvalues are in general related to a pair of primitive solutions which may be written as $v_R(t) \pm v_I(t)$ that directly specify a basic pair of complementary dynamic eigenvalues. There are relationships proven in [1] between the real and intrinsic components, $v_R(t)$ and $v_I(t)$, that improve the understanding and representation of dynamic eigenvalues which are shown next.

$$v(t) = v_R(t) \pm v_I(t) = -\frac{\dot{v}_I(t)}{2\omega_{01}(t)v_I(t)} \pm v_I(t)$$

$$v_I(t) = \frac{1}{p(t)} \; ; \; \sigma_0(t) + \frac{\dot{p}(t)}{2p(t)} = \sigma_0(t) - \frac{\dot{v}_I(t)}{2v_I(t)} \quad (26)$$

$$\lambda(t) = \sigma_0(t) + \omega_{01}(t)v_R(t) \pm \omega_{01}(t)v_I(t)$$

Using the primitive RCE solutions yields core solutions that are reminiscent of the typical solutions given for time invariant systems. Specifically, we may write,

$$x_1(t) = Ae^{\int \lambda(t)dt} \; ; \; \lambda(t) = \sigma_0(t) - \frac{1}{2}\frac{\dot{v}_I(t)}{v_I(t)} \pm \omega_{01}(t)v_I(t)$$

$$x_1(t) = Ae^{\varphi_0(t)}\frac{1}{\sqrt{v_I(t)}}e^{\pm\varphi_I(t)} = \frac{A}{\sqrt{v_I(t)}}e^{\varphi_0(t)\pm\varphi_I(t)} \quad (27)$$

$$\varphi_0(t) = \int \sigma_0(t)dt \; ; \; \varphi_I(t) = \int \omega_{01}(t)v_I(t)dt$$

Note that the primitive RCE solutions spawn the "usual" pair of exponentials in the time domain for the core solutions; however, each is divided by the square root of the RCE intrinsic solution. Observe that complex conjugate primitive RCE solutions spawn complex conjugate exponentials in the time domain. However, the primitive solutions, v(t) shown in (26), for the RCE generate a family of general solutions for the RCE [1], analogous to (24) and (25), applicable to any LTV system. Thus, the general solution for the dynamic eigenvalues is given by,

$$\lambda(t) = \sigma_f(t) + \omega_f(t)\begin{cases} \tanh(\int \omega_f(t)dt - K) \\ \coth(\int \omega_f(t)dt - K) \end{cases}, v_I(t) \; real$$

$$\lambda(t) = \sigma_f(t) + \omega_f(t)\begin{cases} -\tan(\int \omega_f(t)dt - K) \\ \cot(\int \omega_f(t)dt - K) \end{cases} \quad (28)$$

$$\sigma_f(t) = \sigma_0(t) - \frac{1}{2}\frac{\dot{v}_I(t)}{v_I(t)} \; ; \; \begin{array}{l} \omega_f(t) = \omega_{01}(t)v_I(t) \\ \omega_{fm}(t) = \omega_{01}(t)v_{\text{Im}}(t) \end{array}$$

The trigonometric solutions correspond to complex conjugate RCE solutions, where $v_I(t) = jv_{\text{Im}}(t)$. These results provide powerful insight into the solution of LTV systems. Observe that integrating and exponentiating the dynamic eigenvalues produces the time domain solutions for the LTV systems to within a scale factor, A. For the case of real dynamic eigenvalues, using (28), we have a time domain solution given by y(t) where,

$$y(t) = \frac{A}{\sqrt{v_I(t)}}e^{\varphi_0(t)}\begin{cases} \cosh(\varphi_f(t) - K) \\ \sinh(\varphi_f(t) - K) \end{cases} \quad (29)$$

For the case of complex dynamic eigenvalues we have a time domain solution given by,

$$y(t) = \frac{A}{\sqrt{v_{\text{Im}}(t)}}e^{\varphi_0(t)}\begin{cases} \cos(\varphi_{fm}(t) - K) \\ \sin(\varphi_{fm}(t) - K) \end{cases} \quad (30)$$

The results just shown provide the general solution form for all solutions of linear second order systems with their dependence on the intrinsic primitive solution of the RCE. It should be noted that the boundary condition for the RCE, set by the value of K, determines the time domain solution up to a scale factor. On the other hand, the two initial conditions that specify the solution in the time domain, typically y(0) and $\dot{y}(0)$, may be used to find the implicit initial condition pertaining to the RCE solution. The ratio, $\dot{y}(0)/y(0)$, equals $\lambda(0)$ which will specify the relevant value of K for the associated RCE solution.

### IV. Examples

The analysis presented thus far requires serious thought to appreciate its implementation and connection to classic theory. To that end, let us consider some LTI and LTV system examples regarding the solution method offered here.

In each case we begin with a focus on a state matrix for some system. We start with the LTI matrix below, with its associated parameters and matrices.

$$A = \begin{bmatrix} 3 & 1 \\ 2 & 2 \end{bmatrix}; \begin{cases} \alpha = 1/2 \\ \sigma_0 = 5/2 \end{cases} A_0 = \begin{bmatrix} \sigma_0 & \omega_{01} \\ \omega_{02} & \sigma_0 \end{bmatrix}$$

$$\omega_{01} = 1 ; \omega_{02} = 2 + 1(\tfrac{1}{2})^2 = \tfrac{9}{4} ; \dot{v} = -v^2 + \tfrac{9}{4} \quad (31)$$

$$v_p = \sqrt{\tfrac{9/4}{1}} = \pm\tfrac{3}{2} \Rightarrow v_I = \pm\tfrac{3}{2}, v_R = 0, p = \pm\tfrac{2}{3}$$

Note that the first step of transformation yields $A_0$ in modulation form, providing the parameters, $\omega_{01}$ and $\omega_{02}$, which sets up the reduced Riccati characteristic equation for v whose solution will yield the eigenvalues by translation with $\sigma_0$. The reduced Riccati equation is easily solved for constant primitive solutions, $v_p$, yielding their constituent parts, $v_R$ and $v_I$. Continuing, we have,

$$P = \begin{bmatrix} 1 & 0 \\ 0 & 2/3 \end{bmatrix}; A_f = \begin{bmatrix} 5/2 & 3/2 \\ 3/2 & 5/2 \end{bmatrix}; A_{fd} = \begin{bmatrix} 4 & 0 \\ 0 & 1 \end{bmatrix}$$

$$\lambda_{1,2} = \sigma_0 + \omega_{01}(v_R \pm v_I) = \tfrac{5}{2} + 1(0 \pm \tfrac{3}{2}) = 4, 1$$

$$V = \begin{bmatrix} 1 & 1 \\ -\tfrac{1}{2} + \tfrac{1}{2/3} & -\tfrac{1}{2} - \tfrac{1}{2/3} \end{bmatrix} = \begin{bmatrix} 1 & 1 \\ 1 & -2 \end{bmatrix} \quad (32)$$

$$M^{-1}(t) = \begin{bmatrix} 1 & 1 \\ 1 & -2 \end{bmatrix}\begin{bmatrix} e^{4t} & 0 \\ 0 & e^t \end{bmatrix} ; M(0) = \tfrac{1}{3}\begin{bmatrix} 2 & 1 \\ 1 & -1 \end{bmatrix}$$

$$\phi(t) = \tfrac{1}{3}\begin{bmatrix} e^t + 2e^{4t} & e^{4t} - e^t \\ 2e^{4t} - 2e^t & 2e^t + e^{4t} \end{bmatrix}$$

$$y(t) = A_1 e^{4t} + A_2 e^t$$

In this LTI case V(t) is a constant matrix, as expected, with easily found eigenvectors and associated eigenvalues. The state transition matrix is that expected from classical analysis. The time domain solution, y(t), is shown as the usually observed combinations of the core exponentials, $e^{4t}$ and $e^t$, as are the elements of the state transition matrix.

The second example covers a scenario with repeated eigenvalues in the classic case. Observe that the RCE has only a single constant solution–namely, 0. However, allowing time-varying solutions, we get the expected pair of solutions. Following the methodology here we get a straightforward application of the steps.

$$A = \begin{bmatrix} 1 & 1 \\ -1 & 3 \end{bmatrix}; \alpha = -1, \sigma_0 = 2$$

$$\omega_{01} = 1, \omega_{02} = -1 + 1(-1)^2 = 0 ; \dot{v} = -v^2 \quad (33)$$

$$v_p = \tfrac{1}{t}, 0 = \tfrac{1}{2t} \pm \tfrac{1}{2t} = v_R(t) \pm v_I(t)$$

Notice that the primitive solution for the RCE includes no constant of integration and that $v_R(t)$ is related to $v_I(t)$ in the correct way shown in (26). Since p(t) and $v_I(t)$ are reciprocals of one another, p(t) = 2t and q(t) = 1, which sets P(t). Continuing, we have,

$$A_f(t) = \begin{bmatrix} 2+\tfrac{1}{2t} & \tfrac{1}{2t} \\ \tfrac{1}{2t} & 2+\tfrac{1}{2t} \end{bmatrix}; A_{fd}(t) = \begin{bmatrix} 2+\tfrac{1}{t} & 0 \\ 0 & 2 \end{bmatrix}$$

$$\lambda_{1,2} = 2 + 1(\tfrac{1}{2t} \pm \tfrac{1}{2t}) = 2+\tfrac{1}{t}, 2$$

$$V(t) = \begin{bmatrix} 1 & 1 \\ \tfrac{1}{t}+1 & 1 \end{bmatrix}; V^{-1}(t) = t\begin{bmatrix} -1 & 1 \\ 1+\tfrac{1}{t} & -1 \end{bmatrix} \quad (34)$$

$$M^{-1}(t) = V(t)\begin{bmatrix} te^{2t} & 0 \\ 0 & e^{2t} \end{bmatrix} = \begin{bmatrix} te^{2t} & e^{2t} \\ (1+t)e^{2t} & e^{2t} \end{bmatrix}$$

$$\phi(t) = \begin{bmatrix} (1-t)e^{2t} & te^{2t} \\ -te^{2t} & (1+t)e^{2t} \end{bmatrix}$$

$$y(t) = A_1 te^{2t} + A_2 e^{2t}$$

Here we observe, unlike in classical analysis, that the eigenvalues are distinct; however, the associated core modes are $e^{2t}$ and $te^{2t}$ as expected. Coupled with the eigenvectors shown, we obtain the correct solution vectors for the homogeneous system whose state transition matrix is given by $\phi(t)$. The matrix $M_{fd}(t) = V^{-1}(t)$ diagonalizes the state matrix, isolating the distinct eigenvalues. The time domain solution allows for all possible linear combinations of the core solutions.

It is instructive to consider a case with complex eigenvalues. This is done below, starting with the LTI case.

$$A(t) = \begin{bmatrix} 1 & 1 \\ -5 & -3 \end{bmatrix}; \begin{cases} \sigma_0 = -1, \alpha = 2 \\ \omega_{01} = 1, \omega_{02} = -1 \end{cases}$$

$$A_0 = \begin{bmatrix} -1 & 1 \\ -1 & -1 \end{bmatrix}; \dot{v} = -v^2 - 1 ; v_p = \pm j \quad (35)$$

$$p(t) = -j ; q = 0 ; P(t) = \begin{bmatrix} 1 & 0 \\ 0 & -j \end{bmatrix}$$

$$A_f = \begin{bmatrix} -1 & j \\ j & -1 \end{bmatrix}; \lambda_1 = -1+j ; \lambda_2 = -1-j$$

Now the fundamental matrix is given by,

$$M^{-1}(t) = \begin{bmatrix} 1 & 1 \\ -2+j & -2-j \end{bmatrix} \begin{bmatrix} e^{(-1+j)t} & 0 \\ 0 & e^{(-1-j)t} \end{bmatrix} \quad (36)$$

Normalizing the fundamental matrix yields the state transition matrix, with the accompanying system output also shown.

$$\begin{bmatrix} 2e^{-t}\sin(t)+e^{-t}\cos(t) & e^{-t}\sin(t) \\ -5e^{-t}\sin(t) & -2e^{-t}\sin(t)+e^{-t}\cos(t) \end{bmatrix} \quad (37)$$

$$y(t) = e^{-t}(A_1 e^{jt} + A_2 e^{-jt})$$

The primitive solutions for the characteristic RCE are imaginary. Therefore, the eigenvectors are complex, as are the core modes. However, post multiplication of the fundamental matrix, $M^{-1}(t)$ by $M(0)$ produces the purely real state transition matrix. Using the complex eigenvalues, we obtain the expected time domain solution, $y(t)$, where $A_1$ and $A_2$ are complex conjugates that allow the full range of boundary conditions.

There is an interesting alternate way to process the system above in light of the work here. Note that the two primitive solutions to the RCE are $j$ and $j - 2/p(t) = -j$. However, the RCE also has real solutions, which are related to the primitive solutions in the way shown earlier. Furthermore, the complementary pair of these real solutions, shown in (28), may be directly used to find the eigenvector matrix. In this case we have,

$$v_p = 0 \pm j \; ; \; v_I = j = jv_{Im} \Rightarrow v_{Im} = 1$$

$$v_{1,2}(t) = \begin{cases} -1 \cdot \tan(\int 1 dt) = -\tan(t) \\ 1 \cdot \cot(\int 1 dt) = \cot(t) \end{cases} \quad (38)$$

$$V(t) = \begin{bmatrix} 1 & 1 \\ -2-\tan(t) & -2+\cot(t) \end{bmatrix}$$

The complementary pair of solutions also directly yields the dynamic eigenvalues, allowing a quick path to finding the fundamental matrix and the state transition matrix. We begin with the dynamic eigenvalues, exploiting $\omega_{01}(t) = 1$ and $\sigma_0(t) = -1$, which then allows the determination of the fundamental matrix.

$$\lambda_{1,2}(t) = \sigma_0(t) + \omega_{01}(t)v_{1,2}(t) = \begin{cases} -1-\tan(t) \\ -1+\cot(t) \end{cases}$$

$$M^{-1}(t) = V(t) \begin{bmatrix} e^{-t}\cos(t) & 0 \\ 0 & e^{-t}\sin(t) \end{bmatrix} \quad (39)$$

Continuing with this multiplication and normalizing to produce the identity matrix at t equals 0 yields the state transition matrix, which is identical to that shown in (37). The system output, using (25), is the expected real function, $Ae^{-t}\cos(t-K)$, where the dynamic eigenvalues have been exploited to directly provide all intuitive real solutions.

It may now be shown for general LTV system cases that we may appreciate this approach for determining eigenvalues and eigenvectors in a completely consistent way. The next example highlights this fact as well as several other important features of the proposed approach. Starting with the given state matrix, A(t), we can see the effect of the first transformation, $M_0(t)$, exploiting the values of $\sigma_0(t)$ and $\alpha(t)$.

$$A(t) = \begin{bmatrix} \dfrac{t}{t+1} & 1 \\ \dfrac{-5}{4(t+1)^2} & 1+\dfrac{1}{t+1} \end{bmatrix} \begin{cases} \sigma_0(t)=1, \alpha(t)=\dfrac{-1}{t+1} \\ \omega_{02}(t)=\dfrac{3}{4(t+1)^2} \end{cases}$$

$$A_0(t) = \begin{bmatrix} 1 & 1 \\ \dfrac{3}{4(t+1)^2} & 1 \end{bmatrix} \; ; \; \dot{v}(t) = -v^2(t) + \dfrac{3}{4(t+1)^2} \quad (40)$$

$$v_{1,2}(t) = \dfrac{3}{2(t+1)} \, , \, \dfrac{-1}{2(t+1)} = \dfrac{1}{2(t+1)} \pm \dfrac{1}{t+1}$$

Note that $A_0(t)$ is in the modulation form, having an equal pair of diagonal components, and the 1-2 element of the A(t) matrix is unchanged. The off-diagonal elements of $A_0(t)$ spawn the reduced RCE for $v(t)$, with a pair of primitive solutions shown as well. Observe that in this LTV case, $v_R(t)$ is no longer zero as in the LTI case. Knowing the two primitive solutions is sufficient to solve the system. We may directly find the dynamic eigenvalues and the fundamental matrix.

$$\begin{cases} \lambda_1(t) = 1 + \dfrac{3}{2(t+1)} \\ \lambda_2(t) = 1 - \dfrac{1}{2(t+1)} \end{cases} \; ; \; V(t) = \begin{bmatrix} 1 & 1 \\ \dfrac{5}{2(t+1)} & \dfrac{1}{2(t+1)} \end{bmatrix}$$

$$M^{-1}(t) = V(t) \begin{bmatrix} (t+1)^{\frac{3}{2}} e^t & 0 \\ 0 & \dfrac{1}{\sqrt{t+1}} e^t \end{bmatrix} \quad (41)$$

Continuing, we find $\phi(t)$ and y(t).

$$\phi(t) = \dfrac{e^t}{4\sqrt{t+1}} \begin{bmatrix} 5-(t+1)^2 & 2(t+1)^2 - 2 \\ \dfrac{5}{2(t+1)} - \dfrac{5(t+1)}{2} & 5(t+1) - \dfrac{1}{t+1} \end{bmatrix} \quad (42)$$

$$y(t) = \dfrac{1}{\sqrt{t+1}} e^t (A_1(t+1)^2 + A_2)$$

These results are computed in the same way as the time invariant cases. Observe how the time domain solution, y(t),

captures the dynamics of the first row of φ(t), corresponding to the reference $x_1(t)$ state variable. Looking at the eigenvector matrix, V(t), we see how the $x_2(t)$ state variable is related to $x_1(t)$ scaled by $1/(t+1)$.

This was an example of a system with real eigenvalues. The next example below relates to a LTV system with complex dynamic eigenvalues. Note how the time varying imaginary part spawns a real component, even if $\sigma_0(t)$ were zero, which is unique to LTV systems.

$$A(t) = \begin{bmatrix} 1+\frac{1}{t} & 1 \\ -\frac{1}{t^4} & 1-\frac{1}{t} \end{bmatrix} \begin{cases} \sigma_0(t) = 1 \; ; \; \alpha(t) = \frac{1}{t} \\ \omega_{01}(t) = 1 \; ; \; \omega_{02}(t) = -\frac{1}{t^4} \end{cases}$$

$$A_0(t) = \begin{bmatrix} 1 & 1 \\ -\frac{1}{t^4} & 1 \end{bmatrix} \begin{cases} \dot{v}(t) = -v^2(t) - \frac{1}{t^4} \\ v_R(t) \pm v_I(t) = \frac{1}{t} \pm j\frac{1}{t^2} \end{cases} \quad (43)$$

As before, knowing the two primitive solutions is sufficient to solve the system. Again, we may directly find the dynamic eigenvalues and the eigenvector and fundamental matrices.

$$\lambda_{1,2}(t) = 1 + \frac{1}{t} \pm j\frac{1}{t^2} \; ; \; V(t) = \begin{bmatrix} 1 & 1 \\ j\frac{1}{t^2} & -j\frac{1}{t^2} \end{bmatrix}$$

$$M^{-1}(t) = V(t) \begin{bmatrix} te^{t-j/t} & 0 \\ 0 & te^{t+j/t} \end{bmatrix} = \begin{bmatrix} te^{t-j/t} & te^{t+j/t} \\ j\frac{1}{t}e^{t-j/t} & -j\frac{1}{t}e^{t+j/t} \end{bmatrix} \quad (44)$$

This qualifies as a fundamental matrix, since the respective column vectors are each solutions of the homogeneous system with A(t) as its state matrix. However, all linear combinations of these vectors will provide solutions that have $x_1(0) = 0$ and $x_2(0) = \infty$. Consequently, the state transition matrix cannot be found with the identity matrix as its value at t = 0. In order to find a state transition matrix that allows boundary conditions to drive solutions in the usual way, it must be specified to be the identity matrix at a time other than t = 0. This is the kind of problem that can show up in LTV systems, where the state matrix becomes infinite at finite time as in this case. Unbounded finite time solutions in the time domain may only result from unbounded elements in the state matrix as discussed in [1].

Continuing with this example, we may find complementary solutions to the RCE that are purely real as in the LTI example earlier. As before, we use the primitive solution to compute the new solutions to the Riccati equation, and the corresponding system solutions. Noting that $v_{Im}(t) = 1/t^2$, we may write,

$$v_{1,2}(t) = \frac{1}{t} - \frac{1}{t^2}\tan(-\frac{1}{t}) \; , \; \frac{1}{t} + \frac{1}{t^2}\cot(-\frac{1}{t})$$

$$\lambda_{1,2}(t) = 1 + \frac{1}{t} + \frac{1}{t^2}\tan(\frac{1}{t}) \; , \; 1 + \frac{1}{t} - \frac{1}{t^2}\cot(\frac{1}{t}) \quad (45)$$

Continuing, we have,

$$V(t) = \begin{bmatrix} 1 & 1 \\ \frac{1}{t^2}\tan(\frac{1}{t}) & -\frac{1}{t^2}\cot(\frac{1}{t}) \end{bmatrix}$$

$$M^{-1}(t) = V(t) \begin{bmatrix} t\cos(\frac{1}{t})e^t & 0 \\ 0 & t\sin(\frac{1}{t})e^t \end{bmatrix} \quad (46)$$

$$= \begin{bmatrix} t\cos(\frac{1}{t})e^t & t\sin(\frac{1}{t})e^t \\ \frac{1}{t}\sin(\frac{1}{t})e^t & -\frac{1}{t}\cos(\frac{1}{t})e^t \end{bmatrix}$$

We now have a purely real fundamental matrix having the same singularities at t = 0 as in the complex dynamic eigenvalues case.

For completeness and an interesting example of Floquet theory, consider this example from [28] of a periodic LTV system. The system state matrix, A(t), given by,

$$A(t) = \begin{bmatrix} -1 + \frac{3}{2}\sin^2(t) & -1 - \frac{3}{2}\sin(t)\cos(t) \\ 1 - \frac{3}{2}\sin(t)\cos(t) & -1 + \frac{3}{2}\cos^2(t) \end{bmatrix} \quad (47)$$

Proceeding with the analysis, $\sigma_0(t)$ is easily found to be -1/4; however, there is some clutter in the other parameters.

$$\omega_{01}(t) = -1 - \frac{3}{2}\sin(t)\cos(t)$$
$$\omega_{02}(t) = 1 - \frac{3}{2}\sin(t)\cos(t) + \omega_{01}(t)\alpha^2(t) + \dot{\alpha}(t) \quad (48)$$
$$\alpha(t) = \frac{\frac{3}{2}(\cos^2(t) - \sin^2(t))}{2(1 + \frac{3}{2}\sin(t)\cos(t))}$$

The associated Riccati characteristic equation yields the complementary solutions given below, where an effort has been made to reduce the algebraic clutter.

$$\dot{v}(t) = (1 + \frac{3}{2}\sin(t)\cos(t))v^2(t) + \omega_{02}(t)$$
$$v_1(t) = \alpha(t) + \tan(t) \; ; \; v_2(t) = \alpha(t) - \cot(t) \quad (49)$$

Armed with these results we may find the dynamic eigenvalues.

$$\lambda_1(t) = -\tfrac{1}{4} + \omega_{01}(t)(\alpha(t)+\tan(t)) = -1-\tan(t)$$
$$\lambda_2(t) = -\tfrac{1}{4} + \omega_{01}(t)(\alpha(t)-\cot(t)) = \tfrac{1}{2}-\cot(t) \quad (50)$$

Continuing, we may find the eigenvector, fundamental, and state transition matrices.

$$V(t) = \begin{bmatrix} 1 & 1 \\ \tan(t) & -\cot(t) \end{bmatrix}$$
$$M^{-1}(t) = V(t)\begin{bmatrix} \cos(t)e^{-t} & 0 \\ 0 & \sin(t)e^{1/2} \end{bmatrix} \quad (51)$$
$$\phi(t) = \begin{bmatrix} \cos(t) & -\sin(t) \\ \sin(t) & \cos(t) \end{bmatrix}\begin{bmatrix} e^{-t} & 0 \\ 0 & e^{1/2} \end{bmatrix}$$

Observe that one of the dynamic eigenvalues in (50) is clearly unstable in the classic sense since its time domain solution is unbounded. In [28], from where the example is taken and elsewhere, eigenvalues are computed naively as if the state matrix were constant, yielding complex conjugates with a negative real part. Readers are then warned that the apparently stable system is in fact unstable. The analysis here shows that a more sophisticated look clearly reveals the system behavior.

**V. Discussion**

Now we have seen the full scope of the analysis presented here regarding systems and their Riccati characteristic equation solutions. Given the tremendous range of systems available in the LTV category, many more cases than those presented here are possible. Nevertheless, no case fails to meet the framework presented here to the author's knowledge. As shown in [1], the Bessel equation, the quantum harmonic oscillator equation, and the Matthieu equation all exhibit solutions completely consistent with the paradigm of this work. In fact, the analysis here offers new perspectives on the solutions of these well studied engineering systems.

Given that level of completeness, it only remains to summarize and consider possible extensions. Thus far, it has been shown that every second order LTV system has an associated RCE with a primitive pair of solutions that generate an associated continuum of solutions that can be used to meet all boundary conditions. These solutions with an offset and scaling are the dynamic eigenvalues for the system. Solutions in the time domain are had by integrating and exponentiating the dynamic eigenvalues. Owing to a spectrum invariant transformation depending on the RCE solutions and parameters of the state matrix, the state matrix may be canonically diagonalized, having the dynamic eigenvalues on the diagonal. Then a general simple form of the fundamental matrix for the system may be written, incorporating the product of a simple dynamic eigenvector matrix and the exponential of the diagonalized state matrix. Naturally, the state transition matrix is immediately available. As shown in section IV, all of this analysis is completely consistent with that currently used for LTI systems, which are in this context just a natural subset of general LTV systems.

The understanding of the general form of solutions for the RCE given in [1] allows an elegant generalization of classic time domain solutions for LTI systems to be given for LTV systems. A series of examples has fully articulated the use of the proposed methodology for both LTI and LTV systems, including periodic LTV systems. The remaining limitation to what has been proposed here is its restriction to second order systems. A simple extension occurs when a higher order system may be written as a cascade of second order systems, in which case the current method may be applied to each subsystem. Clearly one or more of the subsystems may be first order without issue. The problem becomes difficult when the system state matrix is no longer block diagonal or at least triangular. The work in [6] offers a systematic approach for this, but offers no proof that the operation is spectrum invariant. The problem is compounded in the most general case by the fact that algorithms that can triangularize a large matrix are not guaranteed to group pairs of dynamic eigenvalues in meaningful ways. The extension of the work here to general higher order systems will be valuable. Nevertheless, is seems clear that the methodology presented here can provide important insight to researchers, practitioners, and teachers of linear system theory.

**Appendix:**

Suppose that P(t) is used to transform the state matrix, $A_0(t)$, as given in (9), with the constraint on p(t) given in (13). Then we have,

$$P(t) = \begin{bmatrix} 1 & 0 \\ -q(t) & p(t) \end{bmatrix}; q(t) = \frac{\dot{p}(t)}{2\omega_{01}(t)} = v(t)p(t)-1$$

$$A_f(t) = P(t)\begin{bmatrix} \sigma_0(t) & \omega_{01}(t) \\ \omega_{02}(t) & \sigma_0(t) \end{bmatrix}P^{-1}(t) + \dot{P}(t)P^{-1}(t)$$

$$= \begin{bmatrix} \sigma_0(t) + \dfrac{\dot{p}(t)}{2p(t)} & \dfrac{\omega_{01}(t)}{p(t)} \\ p(t)\omega_{02}(t)+\beta(t) & \sigma_0(t) + \dfrac{\dot{p}(t)}{2p(t)} \end{bmatrix} \quad (A1)$$

$$\beta(t) = -q^2(t)\frac{\omega_{01}(t)}{p(t)} + \frac{\dot{p}(t)}{p(t)}q(t) - \dot{q}(t)$$

Observing that $\dot{p}(t)/p(t) = 2\omega_{01}(t)q(t)/p(t)$, then β(t) is given by,

$$\beta(t) = -q^2(t)\frac{\omega_{01}(t)}{p(t)} + 2\omega_{01}(t)\frac{q^2(t)}{p(t)} - \dot{q}(t)$$
$$= \omega_{01}(t)\frac{q^2(t)}{p(t)} - \dot{q}(t) \quad (A2)$$

Now given that $q(t) = v(t)p(t) - 1$, as shown in (A1), we may rewrite $\beta(t)$, yielding the final form for the 2-1 element of $A_f(t)$.

$$\beta(t) = \frac{\omega_{01}(t)}{p(t)}(v(t)p(t)-1)^2 - (p(t)\dot{v}(t) + \dot{p}(t)v(t))$$
$$= \frac{\omega_{01}(t)}{p(t)} - p(t)[\dot{v}(t) + \omega_{01}(t)v^2(t)] \quad (A3)$$
$$A_{f-21}(t) = \frac{\omega_{01}(t)}{p(t)} + p(t)[\omega_{02}(t) - \dot{v}(t) - \omega_{01}(t)v^2(t)]$$

The term in brackets in the last line above equals zero since $v(t)$ satisfies the RCE by assumption. Hence, the 2-1 element of $A_f(t)$ equals its 1-2 element, $\omega_{01}(t)/p(t)$, and the state matrix assumes the symmetric form. Then, by defining $\sigma_f(t) = \sigma_0(t) + \dot{p}(t)/2p(t)$ and $\omega_f(t) = \omega_{01}(t)/p(t)$, we have the results in (12) and (14).